\documentclass[12pt,preprint]{aastex}

\usepackage{mathrsfs,graphicx}

\begin{document}
\title{Early Quintessence in Light of WMAP}

\author{Robert R. Caldwell}
\affil{Dept. of Physics and Astronomy, Dartmouth College, 
6127 Wilder Laboratory, Hanover, NH 03755}

\author{Michael Doran}
\affil{Dept. of Physics and Astronomy, Dartmouth College, 
6127 Wilder Laboratory, Hanover, NH 03755}

\author{Christian M. M{\"{u}}ller}
\affil{Institut f{\"{u}}r Theoretische Physik, Philosophenweg 16, 
69120 Heidelberg, Germany}

\author{Gregor Sch{\"{a}}fer}
\affil{Institut f{\"{u}}r Theoretische Physik, Philosophenweg 16, 
69120 Heidelberg, Germany}

\author{Christof Wetterich}
\affil{Institut f{\"{u}}r Theoretische Physik, Philosophenweg 16, 
69120 Heidelberg, Germany}

\date{\today}

\newcommand{\omegals}{\Omega^{(\rm ls)}_{\rm q}}
\newcommand{\omegasf}{\Omega^{(\rm sf)}_{\rm q}}
\newcommand{\wbar}{\overline{w}_{\rm q}}
\newcommand{\w}{w_{\rm q}}
\newcommand{\omegaq}{\Omega_{\rm q}}


\begin{abstract}

We examine the cosmic microwave background (CMB) anisotropy for signatures of early
quintessence dark energy -- a non-negligible quintessence energy density during the
recombination and structure formation eras. Only very recently does the quintessence
overtake the dark matter and push the expansion into overdrive. Because the
presence of early quintessence exerts an influence on the clustering of dark matter
and the baryon-photon fluid, we may expect to find trace signals in the CMB and the
mass fluctuation power spectrum. In detail, we demonstrate that suppressed
clustering power on small length-scales, as suggested by the combined Wilkinson
Microwave Anisotropy Probe (WMAP) / CMB / large scale structure data set, is
characteristic of early quintessence. We identify a set of concordant models, and
map out directions for further investigation of early quintessence.

\end{abstract}

\maketitle


There exists compelling evidence that the energy density of the Universe is
dominated by dark energy. The evidence grows stronger with each successive
experiment and observation of cosmic evolution and structure, as boldly reinforced
by the recent high precision measurement of the cosmic microwave background (CMB)
fluctuations by WMAP
\citep{Bennett:2003bz,Spergel:2003cb,Kogut:2003et,Hinshaw:2003ex,Verde:2003ey,Page:2003fa}.
And yet, the nature of the dark energy remains elusive.  A cosmological constant
($\Lambda$), providing a simple phenomenological fix in the absence of better
information, is consistent with current data including the latest WMAP results. 
Lessons from particle physics and cosmology, however, suggest a more attractive
solution in the form of a dynamical dark energy
\citep{Wetterich:fm,Ratra:1987rm,Peebles:1987ek,Caldwell:1997ii} that continues to
evolve in the present epoch  --- quintessence. 

In the  quintessence scenario the dark energy becomes dominant only at late
times, as required for cosmic acceleration. However, the late appearance of the
quintessence may not be the whole story.  Scalar field models of quintessence
with global attractor solutions
\citep{Wetterich:fm,Ratra:1987rm,Zlatev:1998tr,Steinhardt:nw} have been shown
to ``track'' the dominant component of the cosmological fluid. One consequence
is that just after inflation, the universe may contain a non-negligible
fraction of the cosmic energy density. Through subsequent epochs, the
quintessence energy density $\rho_{\rm{q}}$ lags behind the dominant component
of the cosmological fluid with a slowly varying $\omegaq$, and an
equation-of-state $\w \equiv p_{\rm{q}}/\rho_{\rm{q}}$ which is nearly
constant.  The field energy tracks the background until the current epoch, when
the quintessence energy density crosses and overtakes the matter density. A
non-negligible fraction of dark energy at last scattering, $\omegals$, and
during structure formation, $\omegasf$, then arises quite  naturally. From the
observational viewpoint, detection of any trace of ``early quintessence'' would
give us a tremendous clue as to the physics of dark energy.

In this work we concentrate on ``early quintessence'', characterized by
non-negligible values $\omegals,\, \omegasf \lesssim 0.05$. Typical scalar
field models exhibit an exponential form of the scalar potential in the range
of the field relevant for early cosmology,  with special features in the
potential or kinetic term in the range governing the present epoch
\citep{AW,Albrecht,Armendariz,CL}. We describe such models in more detail
below. Our attention is drawn toward these models due to the recent claims of
suppressed power on small scales in the combined WMAP / CMB / large scale
structure data set. We are motivated precisely by the fact that the most
prominent influence of a small amount of early dark energy is a suppression of
the growth of dark matter fluctuations \citep{Doran:2001rw,FJ}.  As we soon
discuss, this influence can help to make the fluctuation amplitude extracted
from galaxy catalogues or the Ly-$\alpha$ forest compatible with a relatively
high amplitude CMB anisotropy.

The effect of early quintessence on the mass fluctuation power spectrum can be
understood simply as a suppression of the growth function for dark matter and
baryonic fluctuations. Just as fluctuation growth is suppressed at late times with
the onset of dark energy, so is the growth of linear modes slowed at early times due
to non-negligible $\omegals,\,\omegasf$. For modes which enter the horizon before
equality, $k > k_{\rm eq}$, the effect is an overall suppression. For $k < k_{\rm eq}$,
the suppression only takes place after the mode enters the horizon.
The consequence is that the smaller scale matter density
perturbations are more suppressed, which ultimately appears as a scale-dependent red {\it
tilt} for the $k < k_{\rm eq}$ modes and a flat suppression for the $k >  k_{\rm eq}$ modes.

We can directly examine the effect on the mass power spectrum by comparing the
$\sigma_8$ values of an early quintessence model with a $\Lambda$ model having the
same amount of present-day dark energy. Fixing the amplitude of the CMB fluctuations
over a range of angular multipoles corresponding to $k \approx (8
\,\textrm{Mpc})^{-1}$, then
\begin{equation}\label{sigma_equation}
\frac{\sigma_8(Q)}{\sigma_8(\Lambda)}=
(a_{eq})^{3 \omegasf/5}(1-\omegaq^{(0)})^{-(1+\tilde w^{-1})/5}
\sqrt{\frac{\tau_0(Q)}{\tau_0(\Lambda)}}.
\end{equation}
The dominant effect is the first factor with $a_{eq} = \Omega_r/\Omega_m \approx
1/3230$. This factor accounts for the slower growth of the cold dark matter
fluctuations. The other kinematical factors involve a suitably-averaged quintessence
equation-of-state in the recent epoch, $\tilde{w}$
\citep{DLSW,Huey:1998se,Perlmutter:1999jt}, and the present conformal time $\tau_0$
for the quintessence and $\Lambda$ models. We emphasize that
eq.(\ref{sigma_equation}) results in a uniform suppression of the cold dark matter
amplitudes for all modes that have entered the horizon since $z_{eq}$.

\begin{figure*}
    \plotone{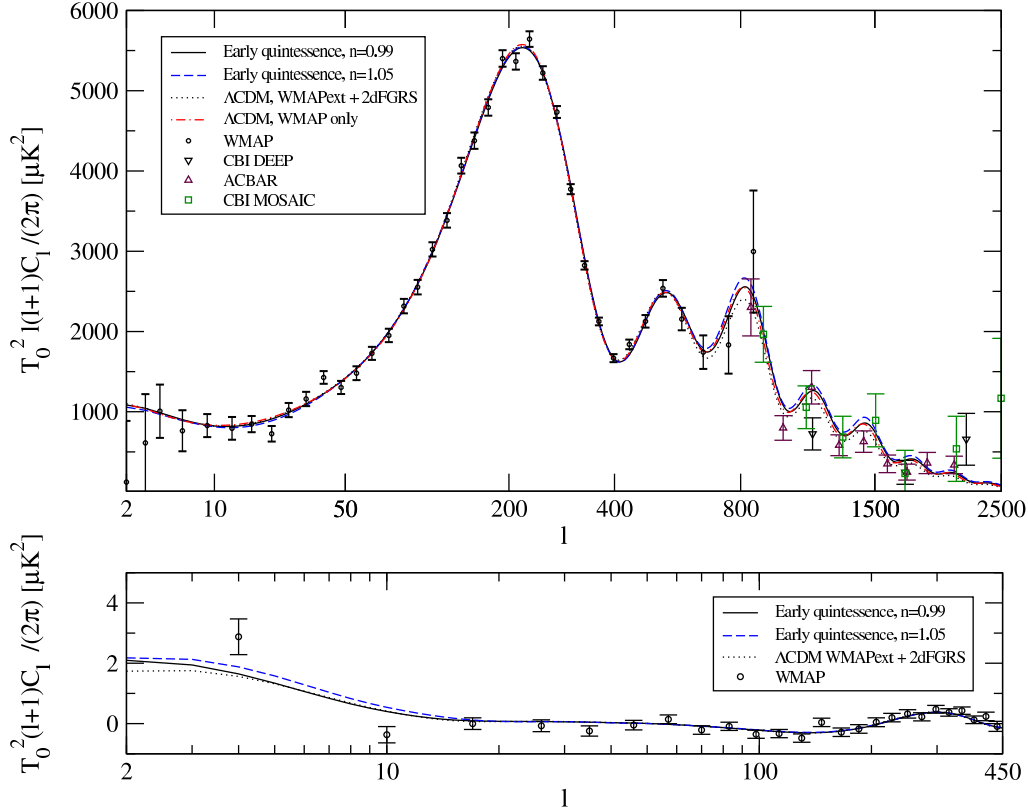}
    \caption{Temperature (TT) and Polarization (TE) as a function of multipole $l$.
        The WMAP data \citep{Hinshaw:2003ex,Kogut:2003et} are plotted alongside
        two early quintessence models with $n_s=0.99$ and $n_s=1.05$ (see
        Table \ref{models} for the other cosmological parameters). For comparison,
        we plot WMAP-normalized spectra for the best fit $\Lambda$CDM model (no 
        Ly-$\alpha$ data) with constant spectral index $n=0.97$ of
        \citep{Spergel:2003cb}, as well as the best fit $\Lambda$CDM model with
        running spectral index $n_s=0.93,\ {\rm d} n_s / {\rm d} \ln k = -0.031$.
        At large $l$ we plot the CBI \citep{CBI_mosaic,CBI_deep} and ACBAR
        \citep{ACBAR} measurements.}
    \label{cmb}
\end{figure*}

Now we turn to consider the implications of the CMB for quintessence. The
temperature anisotropy power spectrum, from the plateau through the first two peaks,
now has been measured with new accuracy. In the context of a spatially-flat
$\Lambda$ model, this would tell us the Hubble constant, $h$, matter and baryon
densities, $\Omega_{\rm m}$ and $\Omega_{\rm b}$, very precisely. For the case of
quintessence, a degeneracy exists amongst these parameters, and the influence of the
equation-of-state can play off the Hubble constant to achieve an otherwise
indistinguishable anisotropy pattern out to small angular scales \citep{Huey:1998se}.
Clearly, the CMB sky is consistent with a small amount of early quintessence in
addition to $\omegaq^{(0)}$ insofar as the angular-diameter distance to the last
scattering surface is preserved. As a means of proof by example, we identify a set
of models in Table~\ref{models} with observationally indistinguishable CMB patterns,
{\it i.e.} identical peak positions, but differing amounts of $\omegals,\,\omegasf$,
shown as Models (A,B) in Figure~\ref{cmb}. Model (C) is WMAP's best fit $\Lambda$CDM
and Model (D) the best fit for an extended data set with $\Lambda$CDM and running
spectral index \citep{Spergel:2003cb}. Our methodology, therefore, is to use the CMB
data to guide our search for compatible quintessence models, rather than carrying
out an exhaustive survey of parameter space. We choose models with present-day
equation-of-state $\w^{(0)} \lesssim -0.9$ so as to focus attention on the early
rather than late quintessence behavior, as compared to a $\Lambda$ model. Because a
significant parameter degeneracy between the primordial scalar spectral index $n_s$
and the optical depth to last scattering $\tau$ persists in the WMAP data, we
explore different combinations of $n_s,\, \tau$.


\begin{table}[h] \caption{\label{models}Models and parameters}
  \begin{tabular}{ccccc}
                 & A & B & C & D \\ \hline
  $\omegasf$     &  0.03 & 0.05  & 0  & 0  \\
  $\omegals$     &  0.03 &  0.05 & 0  & 0  \\ 
  $\w^{(0)}$     & -0.91 & -0.95  & -1   & -1      \\ 
  $n_s$          & 0.99  & 1.05  & 0.97   & 0.93 \\ 
  $h$            & 0.65  & 0.70  & 0.68   & 0.71  \\ 
  $\Omega_m h^2$ & 0.15  & 0.16  & 0.15  & 0.136 \\
  $\Omega_b h^2$ & 0.024 & 0.025  & 0.023 & 0.022 \\
  $\tau$         & 0.17  & 0.26   & 0.1     & 0.17    \\  \hline 
  $\sigma_{8}$   & 0.81  & 0.87 & 0.87 & 0.85 \\
  WMAP: $\chi^2_{\rm eff.}/ \nu$ &1432/1342  & 1432/1342  & 1430/1342 &  1432/1342\\
  CBI-MOSAIC: $\chi^2_{\rm eff.} / \nu$ & 1.1/3  & 1.8/3  & 0.78/3 &  0.34/3\\
  ACBAR: $\chi^2_{\rm eff.} / \nu$ &7.0/7  & 6.4/7  & 6.7/7 &  6.1/7\\
  2dFGRS: $\chi^2_{\rm eff.} / \nu$ & 29/32  & 28/32  & 27/32 &  29/32\\

 \end{tabular}
\end{table}

\begin{figure}
\plotone{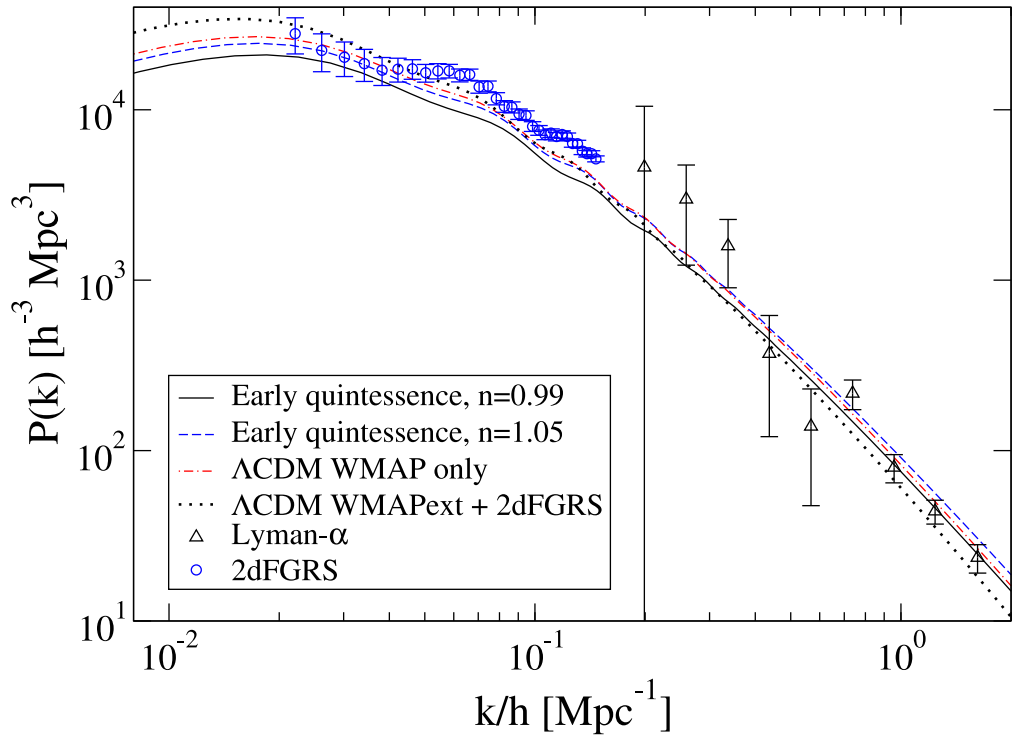}
   \caption{The cold dark matter power spectrum today as a function of $k/h$. We
                plot the linear spectrum for two early quintessence models with
                spectral indices $n_s=0.99$ and $n_s=1.05$ (see Table \ref{models}
                for the other cosmological parameters). Also shown is the best fit
                $\Lambda$ model with running spectral index $n_s=0.93,\ {\rm d}
                n_s / {\rm d} \ln k = -0.031$ of \citep{Spergel:2003cb}, normalized
                to WMAP (no Ly-$\alpha$ data).  The 2dFGRS
                \citep{Percival,Peacock,Verde} and Ly-$\alpha$ \citep{Gnedin,Croft}
                data have been evolved to $z=0$, although we have not convolved our
                theoretical data with the experimental window functions. The galaxy
                power spectrum has a bias compared to theoretical predictions which
                is not included in the figure. }
    \label{power_spectrum}
\end{figure}

In Fig.\ref{power_spectrum} we compare the prediction of our models for the matter
power spectrum with data extracted from galaxy catalogues (e.g. 2dFGRS
\citep{Percival,Verde,Peacock} or the Ly-$\alpha$ forest \citep{Gnedin,Croft}).  In
view of the uncertainties from bias and nonlinearities, the agreement is good for
all models (A-D).  

Increasing the spectral index to $n_s>1$ induces more power for  the fluctuation
spectrum on small scales relative to large. We remark that this enhancement of
small-scale power can be balanced by an increase in $\omegasf$. Typically, for
$\sigma_8$ to remain constant a 10\% increase of $n_s$ is compensated by a 5\%
increase of $\omegasf$. Consequently we find a degeneracy in the $n_s - \omegasf$
plane for $\sigma_8$. (See Fig. 3d of \citep{Doran:2001rw}.)   The degeneracy may be
broken once data for much larger $k$ is included, such as the Ly-$\alpha$ forest.
Whereas $\omegasf$ leads to a uniform decrease of all mass fluctuations with $k/h >
0.064 \, \mbox{Mpc}^{-1}$ by a constant factor, the increase of the small scale
matter fluctuations due to $n_s$ depends on scale $\propto k^{n_s}$. 

An increase of $n_s$ also influences the detailed CMB spectrum in a number of ways. 
First, the spectral index influences the precise location of the first peak in
angular momentum space $l_1$. Parametrizing the location of the peaks as $l_m = l_A
(m-\varphi_m)$ \citep{Hu}, one observes that the shift $\varphi_1$ decreases by 4.7\%
if $n_s$ increases by 10\%. Keeping the well-measured position of $l_1$ fixed
\citep{Page:2003fa}, this results in a decrease of $l_A$ by 5\%. As a consequence,
the location of the second and third peak are shifted by $\Delta l_2 \approx 19$, $
\Delta l_3 \approx 38$ towards smaller $l$. Again, this effect can (partly) be
compensated by an increase of $\omegals$ according to $\varphi_1 \approx [1-0.466
(n_s-1)][0.2604+0.291 \omegals]$ \citep{Doran:2001yw}.  Second, increasing $n_s$
lowers the amplitude ratio between the second and first peak. This can to be
compensated by a larger fraction of baryons $\Omega_b h^2$. Third, larger $n_s$ adds
power to the CMB spectrum at large $l$, or a lower relative power at low $l$.  To
the extent that WMAP and COBE \citep{COBE:dmr4,COBE:2pt4,COBE:rms4} observe a lack of
power on large scales, $l\lesssim 10$, then a blue tilt is beneficial, as a 10\% 
gain in $n_s$ lowers the quadrupole relative to $l=40$ by a factor of $\sim 1.8$,
more in line with observations.

In the extended WMAP parameter analysis, combining CMB with non-CMB cosmological
constraints, the running $k$-dependence in $n_s$ is shown to  lower the matter power
spectrum at $\sigma_8$ and smaller scales, as well as reduce the small-angle CMB
fluctuation power, without touching the region of $l$ measured by WMAP
\citep{Spergel:2003cb}. In the case of early quintessence, the CMB power spectrum on
small angular scales is only mildly lowered. However, the matter fluctuations turn
out to be smaller for a given CMB amplitude. The net effect is a shift of the CMB
power extrapolated from structure formation data towards larger values.  

The models of early quintessence considered in this article can be described
by a scalar field with a non-standard kinetic term evolving in an
exponential potential. The field Lagrangian is
\begin{equation}\label{eqn::leap}
\mathcal{L} =\frac{1}{2} k^2(\varphi) \partial_\mu \varphi \partial^\mu \varphi + M^4 \exp(-\varphi / M).
\end{equation} 
and corresponds to the case of a cosmon field with a leaping kinetic term
\cite{AW}.
Many more models of early quintessence can be cast into the form (\ref{eqn::leap})
by an appropriate rescaling of the scalar field. 
Typically, the function $k^2(\varphi)$ is nearly
constant at early times, leading to the well known attractor dynamics of
quintessence fields with exponential potentials 
\cite{Wetterich:fm,Ratra:1987rm,FJ,Copeland:1997et}. Hence, in the early
Universe one has $\omegaq = \mathcal{N}  k^2(\varphi)$ with $\mathcal{N}=3,4$
for matter and radiation domination, respectively. No tuning of parameters is
required in order to explain why dark energy has been of a similar order of
magnitude as radiation and matter in most cosmological epochs (albeit
substantially smaller in early times, e.g. $\omegaq^{(\rm ls)}$  a few
percent). For a realistic model,  however, $k^2(\varphi)$ must grow in the late
universe. The increased weight of the kinetic term in the Lagrangian due to
this growth leads to a drain of kinetic energy $T$ into potential energy $V$,
effectively ``stopping'' the field. As the pressure of quintessence is given by
$T-V$, this change in $k^2(\varphi)$ leads to negative pressure accelerating
the universe today. 
 For a specific example, the 
 recent increase of
$k^2(\varphi)$ is related to the renormalization group running of the wave
function renomalization of  a dilaton-type model \citep{CL}.  This needs a mild tuning of a
parameter (0.1 \% level) in order for the crossover to quintessence domination to happen now. 
Interestingly, this crossover in the dynamical behavior of the scalar field may also
be observable as a jump in the rate of variation of the fine structure constant.

We have computed the spectra in Figures~\ref{cmb},\,\ref{power_spectrum} using
\mbox{CMBEASY} \citep{Cmbeasy} for a class of ``leaping kinetic term quintessence''
\citep{AW} models with early quintessence.
In these models $k(\varphi)$ makes a relatively rapid jump from a small to a large value at some
approximately chosen value of $\varphi$.
 The main features depend only on
two parameters besides the present fraction of dark energy $\omegaq^{(0)}$ and the
present equation of state $\w^{(0)}$, namely the fraction of dark energy at last
scattering, $\omegals$, and during structure formation, $\omegasf$. In  order to
facilitate comparison with other effects of quintessence -- for example the Hubble
diagram $H(z)$ for supernovae -- we present a useful parametrization of quintessence
\citep{CL} rather than detailed models. For $a> a_{eq}$ and $x \equiv \ln a = - \ln
(1+z)$ we consider a quadratic approximation for the averaged equation-of-state
$(x_{ls} \approx -\ln(1100))$
\begin{eqnarray}
\wbar(x) &=& - \frac{1}{x}\int^0_x dx' \w(x')  \\ \nonumber
         &=& \w^{(0)} + (\wbar^{(ls)}-\w^{(0)}) \frac{x}{x_{ls}} + A x (x-x_{ls}).
\end{eqnarray}
The time-dependent average equation of state $\wbar(x)$ is directly connected to the
time history of the fraction in dark energy $\omegaq (x)$ according to
\begin{equation}
\frac{\omegaq(x)}{1-\omegaq(x)}= \frac{\omegaq^{(0)}(1+a_{eq})}{1-\omegaq^{(0)}}\frac{\exp(-3x \wbar(x))}{1+a_{eq}\exp(-x)}
\end{equation}
which connects $\wbar^{(\rm ls)}$ to $\omegaq^{(\rm ls)}$. The parameter $A$ is
related to the average fraction of dark energy during structure formation ($a_{tr}
\approx 1/3$)
\begin{equation}
\omegasf = \int^{\ln a_{tr}}_{\ln a_{eq}}\frac{ \omegaq(a) \rm{d}\ln a }{\ln(a_{tr}/a_{eq})}.
\end{equation}
The parameters describing our models are  (A): $\wbar^{(\rm ls)}=-0.188$,
$A=-0.0091$; (B): $\wbar^{(\rm ls)}=-0.172$, $A=-0.015$. The Hubble expansion has a
simple expression in terms of $\wbar(x)$
\begin{eqnarray}
H^2(z)&=&H_0^2 \bigg[ \omegaq^{(0)}(1+z)^{3(1+\wbar(x))}+ \nonumber \\ 
 &  &  \Omega_m^{(0)}\Big((1+z)^3+a_{eq}(1+z)^4 \Big) \bigg ].
\end{eqnarray}
Our models (A) and (B) are consistent with all present bounds for $H(z)$, including
type 1a supernovae
\citep{Schmidt:1998ys,Riess:1998cb,Garnavich:1998th,Perlmutter:1998np,Perlmutter:1999jt}.


To summarize, we have demonstrated that models of early quintessence are compatible
with the presently available data for a constant spectral index of primordial density
perturbations. Parameter degeneracies in the angular-diameter distance to last
scattering are consistent with a small abundance of early quintessence. In turn, the
presence of early quintessence results in a differential reduction or scale-dependent
tilt in the spectrum of matter fluctuations on  scales $k < k_{\rm eq}$ and a uniform suppression
of power for scales $k > k_{\rm eq}$, which may have
significant consequences for the interpretation of combined CMB and large scale
structure data. We note that special care must be taken now to interpret large scale
structure observations in the context of early quintessence models. At the other end
of the spectrum, the lack of very large scale power in the CMB can be compensated in
part by increasing both the primordial spectra tilt and increasing the amount of
early quintessence.  We look ahead toward on-going and future tests which afford
tighter measurements of small scale CMB and matter power spectra. A precision
measurement of the position and height of the third peak could be extremely helpful
for determining the fraction of early quintessence.

\begin{acknowledgements}

We thank E. Thommes, P. Butterworth, W. Percival and A. Hamilton for helpful
discussions. R. Caldwell and M. Doran are supported by NSF grant PHY-0099543, C.M.
M{\"{u}}ller and G. Sch{\"{a}}fer are supported by GRK grant 216/3-02.

\end{acknowledgements}


\bibliographystyle{unsrt}

\end{document}